\documentclass{llncs}
\usepackage{times}
\usepackage{courier}
\usepackage{latexsym}
\usepackage{url}
\usepackage{xspace}
\usepackage{amsmath}
\usepackage{amssymb}
\usepackage{microtype}
\usepackage{tabularx}
\usepackage{tikz}
\usepackage{listings}
\newcommand{\mi}[1]{\ensuremath{\mathit{#1}}}

\newcommand{\define}[1]{\emph{#1}}

\newcommand{\la}{\leftarrow}

\newcommand{\tool}[1]{\texttt{#1}\xspace}

\newcommand{\clingo}{\tool{clingo}}
\newcommand{\prolog}{\tool{prolog}}

\newcommand{\setbs}[0]{\ensuremath{\mathcal{BS}}\xspace}
\newcommand{\setkb}[0]{\ensuremath{\mathcal{KB}}\xspace}
\newcommand{\setacc}[0]{\ensuremath{\mathcal{ACC}}\xspace}
\newcommand{\accN}[0]{\ensuremath{\mathrm{ACC}}\xspace}

\newcommand{\bs}[0]{\ensuremath{S}\xspace}

\newcommand{\relout}[4]{\ensuremath{{d_{#1}}({#2,#3,#4})}\xspace}

\newcommand{\contexts}[0]{\ensuremath{\mathsf{C}}\xspace}

\newcommand{\onames}[0]{\ensuremath{\mathsf{O}}\xspace}

\newcommand{\eoc}[0]{\ensuremath{\mathrm{EOC}}\xspace}

\newcommand{\C}[1]{\ensuremath{C_{#1}}\xspace}

\newcommand{\IL}[0]{\ensuremath{\mathcal{IL}}\xspace}
\newcommand{\ls}[0]{\ensuremath{\mathcal{LS}}\xspace}
\newcommand{\outr}[0]{\ensuremath{\mathrm{OR}}\xspace}
\newcommand{\ib}[0]{\ensuremath{\mathrm{ib}}\xspace}

\newcommand{\cuf}[0]{\ensuremath{\mathrm{cu}}\xspace}
\newcommand{\cc}[0]{\ensuremath{\mathrm{cc}}\xspace}
\newcommand{\cname}[0]{\ensuremath{\name}\xspace}

\newcommand{\names}[0]{\ensuremath{\mathcal{N}}\xspace}

\newcommand{\data}[0]{\ensuremath{\mathit{d}}\xspace}
\newcommand{\info}[0]{\ensuremath{\mathsf{i}}\xspace}
\newcommand{\outName}[0]{\ensuremath{\mathsf{o}}\xspace}
\newcommand{\name}[0]{\ensuremath{\mathsf{n}}\xspace}

\newcommand{\head}[1]{\ensuremath{\mi{hd}({#1})}\xspace}
\newcommand{\body}[1]{\ensuremath{\mathrm{bd}({#1})}\xspace}

\newcommand{\kb}[0]{\ensuremath{\mathrm{KB}}\xspace}

\newcommand{\naf}[0]{\ensuremath{\mathrm{not}}\xspace}

\newcommand{\tuple}[1]{\ensuremath{\tupleLeft {#1} \tupleRight}\xspace}
\newcommand{\tupleLeft}[0]{\ensuremath{\langle}\xspace}
\newcommand{\tupleRight}[0]{\ensuremath{\rangle}\xspace}

\newcommand{\aMCSinternal}{aMCS}
\newcommand{\amcs}{\aMCSinternal\xspace}
\newcommand{\amcss}{\aMCSinternal{}s\xspace}

\newcommand{\iec}[0]{i.e.,\xspace}
\newcommand{\cf}[0]{cf.\xspace}
\newcommand{\egc}[0]{e.g.,\xspace}

\newcommand{\bi}[0]{\begin{itemize}}
\newcommand{\ei}[0]{\end{itemize}}
\newcommand{\be}[0]{\begin{enumerate}}
\newcommand{\ee}[0]{\end{enumerate}}

\begin{document}

\title{Stream Packing for\\ Asynchronous Multi-Context Systems using ASP\thanks{This work has been partially supported by the German Research Foundation (DFG) under grants BR-1817/7-1 and FOR 1513.}}

\author{Stefan Ellmauthaler \and J\"org P\"uhrer\institute{Institute of Computer Science,\\ Leipzig University, Germany,\\\{ellmauthaler,puehrer\}@informatik.uni-leipzig.de} }

\maketitle

\begin{abstract}
When a processing unit relies on data from external streams, we may face the problem
that the stream data needs to be rearranged in a way that allows the unit to perform its task(s).
On arrival of new data, we must decide whether there is sufficient information available
to start processing or whether to wait for more data.
Furthermore, we need to ensure that the data meets the input specification of the processing step.
In the case of multiple input streams it is also necessary to coordinate which data from which incoming stream
should form the input of the next process instantiation.
In this work, we propose a declarative approach as an interface between multiple streams and a processing unit.
The idea is to specify via answer-set programming how to arrange incoming data in packages
that are suitable as input for subsequent processing.
Our approach is intended for use in \define{asynchronous multi-context systems} (\amcss),
a recently proposed framework for loose coupling of knowledge representation formalisms that allows for online reasoning in a dynamic environment.
Contexts in \amcss process data streams from external sources and other contexts.%
\end{abstract}

\section{Introduction}%
The omnipresence of smart devices and recent advancements towards a Semantic Web and the Internet of Things
has increased the interest in reasoning over streaming data~(\egc \cite{Brewka2014,EllmauthalerP14,GKL2014,Le-Phuoc2012,BeckDEF15}).
In this work, we take a closer look at \define{asynchronous multi-context systems} (\amcss)~\cite{EllmauthalerP14}.
The \amcs framework allows for loose coupling of knowledge representation formalisms and, at the same time, for online reasoning in a dynamic environment.
Unlike other recent proposals for online multi-context systems~\cite{Brewka2014,GKL2014}, contexts of an \amcs run independently of each other, \iec there are no synchronised time steps in which all contexts exchange information.
Instead, after each evaluation of a context, it sends its results to other contexts and, from the perspective of the receiving side, a context can continuously receive data from other contexts or the outside world in an unpredictable order. 
As the context needs some time for each evaluation and since there might not always be sufficient or the right kind of data in order to start another evaluation,
the incoming data needs to be buffered.
In this work we deal with the problem of how and when to take information from the buffer to hand it over to the context for processing.

In a running example we deal with an 
\amcs with a context that can assign ambulances to an emergency case.
The input buffer of this context can contain information about multiple unassigned cases and multiple available ambulances.
We need to ensure that for every evaluation we only pass on information about a single emergency case (\egc the one with the highest priority) but all information about available rescue units,
in order to meet the input requirements of the knowledge base of the context.
On the other hand, if no information about a case or no ambulance is available yet, the context cannot start, so we need to wait for more data.
We propose to use answer-set programming (ASP)~\cite{MT99,N99,Baral2003}
to declaratively specify when a context may start processing and how
to combine the arrived information into logical packages suitable for the context.
Decisions are based on meta information such as the amount of available data or the source of the information.
Furthermore, by tagging incoming data, \egc also information about time of arrival or the content of the data can be taken into account.
With this approach we give the context to possibility to filter the data it receives.
We define a range of dedicated ASP atoms that can be used in answer-set programs to specify packages
and manipulate data in the input buffer.
In addition, we discuss creating multiple packages at once and how we can exploit ASP optimisation features.
\par 

The remainder of this work is structured as follows:
Next, we will give some background on \amcs in form of a short overview and introduce the example scenario. 
Then, the main approach is introduced in Section~\ref{sec:main} and discussed along various illustrative examples.
Finally, we will conclude in Section~\ref{sec:conclusion}.

\section{Asynchronous Multi-Context Systems}%

We next introduce the \amcs framework, focusing on aspects relevant to this work.
In particular, we describe the asynchronous semantics of \amcss from the perspective of a single context, and refer the interested reader to~\cite{EllmauthalerP14} for a precise characterisation of the semantics of an \amcs as a whole.
Every context in an \amcs is associated with a \emph{logic suite}~\cite{BrewkaEFW11} which can be seen as an abstraction of different KR formalisms.
A logic suite is a triple $\ls=\tuple{\setkb,\setbs, \setacc}$, where $\setkb$ is the set of admissible knowledge bases (KBs) of $\ls$. 
\setbs is the set of possible belief sets of \ls, whose elements are \define{beliefs}.
$\setacc$ is a set of semantics for $\ls$: a semantics for $\ls$ is a function $\accN : \setkb \to 2^{\setbs}$ assigning to each KB a set of acceptable belief sets.
Using a semantics with potentially more than one acceptable belief set allows for modelling non-determinism, where each belief set corresponds to an alternative solution.
We assume a set $\names$ of \define{names} that serve as labels for sensors, contexts, and output streams.
A \define{context} is a pair $C=\tuple{\cname,\ls}$ where $\cname\in\names$ is the name of the context and \ls is a logic suite.

\begin{definition}[\cite{EllmauthalerP14}]
An \amcs (of length $n$ with $m$ output streams) is a pair $M=\tuple{\contexts,\onames}$, where 
$\contexts=\tuple{C_1,\dots,C_n}$ is an $n$-tuple of contexts
and $\onames=\tuple{\outName_1,\dots,\outName_m}$ with $\outName_j\in\names$ for each $1\leq j\leq m$ is a tuple containing the names of the output streams of $M$.
\end{definition}
A context in an \amcs communicates with other contexts and the outside world by means of streams of data.
In particular, every context has an input stream on which information can be written from both external
sources (called sensors) and internal sources (\iec other contexts).
For the data in the communication streams we assume a communication language \IL where every $\info\in\IL$ is an abstract \define{piece of information}.
Both, the data in the input stream of a context and the data in output streams are modelled by \define{information buffers} that are defined in the following.
\begin{definition}[\cite{EllmauthalerP14}]\label{def:data}
A \define{(source-tagged) data set} is a pair $\data=\tuple{\cname,I}$, where $\cname\in\names$ is either a context name or a sensor name, stating the 
\define{source} of \data, and $I\subseteq\IL$ is a set of pieces of information.
An \define{information buffer} is a sequence of data sets\footnote{Note that data sets are called \emph{data packages} in~\cite{EllmauthalerP14}. We changed the name to avoid confusion with packages as introduced later in the paper.}.
\end{definition}

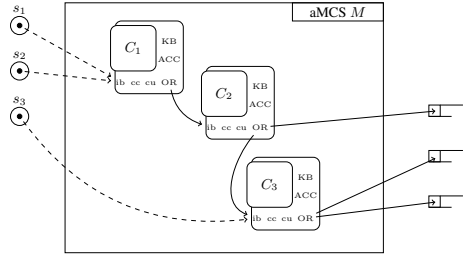
\begin{figure}[t]
  \centering
  \resizebox{0.5\columnwidth}{!}{%
\begin{tikzpicture}
  \tikzstyle{ttext}=[font=\tiny,anchor=base]
  \draw (1,2) rectangle (8,7.5);
  \draw (6,7.1) rectangle  node {\amcs $M$} (8,7.5);
  \draw [fill=black] (0,7) circle (0.05);
  \draw (0,7) circle (0.2) node[label={[label distance=1]$s_1$}] (s1) {};
  \draw [fill=black] (0,6) circle (0.05);
  \draw (0,6) circle (0.2) node[label={[label distance=1]$s_2$}] (s2) {};
  \draw [fill=black] (0,5) circle (0.05);
  \draw (0,5) circle (0.2) node[label={[label distance=1]$s_3$}] (s3) {};
  \draw (9,5) rectangle (9.25,5.25) node [yshift=-3.5] (out1) {};
  \draw (9.25,5.25) -- (9.75,5.25);
  \draw (9,5) -- (9.5,5);

  \draw (9,4) rectangle (9.25,4.25) node [yshift=-3.5] (out2) {};
  \draw (9.25,4.25) -- (9.75,4.25);
  \draw (9,4) -- (9.5,4);

  \draw (9,3) rectangle (9.25,3.25) node [yshift=-3.5] (out3) {};
  \draw (9.25,3.25) -- (9.75,3.25);
  \draw (9,3) -- (9.5,3);
  \draw [rounded corners] (5.1,2.5) rectangle (6.6,4.1);
  \draw [fill=white,rounded corners] (5,3) rectangle node{$\C{3}$} (6,4);
  \node [ttext](kb3) at (6.29,3.6) {$\kb$};
  \node [ttext](acc3) at (6.29,3.2) {$\accN$};
  \node [ttext](ib3) at (5.22,2.7) {$\ib$};
  \node [ttext](cc3) at (5.55,2.7) {$\cc$};
  \node [ttext](cuf3) at (5.88,2.7) {$\cuf$};
  \node [ttext](or3)  at (6.29,2.7) {$\outr$};

  \draw [rounded corners] (2.1,5.5) rectangle (3.6,7.1);
  \draw [fill=white,rounded corners] (2,6) rectangle node{$\C{1}$} (3,7);
  \node [ttext](kb1) at (3.29,6.6) {$\kb$};
  \node [ttext](acc1) at (3.29,6.2) {$\accN$};
  \node [ttext](ib1) at (2.22,5.7) {$\ib$};
  \node [ttext](cc1) at (2.55,5.7) {$\cc$};
  \node [ttext](cuf1) at (2.88,5.7) {$\cuf$};
  \node [ttext](or1)  at (3.29,5.7) {$\outr$};

  \draw [rounded corners] (4.1,4.5) rectangle (5.6,6.1);
  \draw [fill=white,rounded corners] (4,5) rectangle node{$\C{2}$} (5,6);
  \node [ttext](kb2) at (5.29,5.6) {$\kb$};
  \node [ttext](acc2) at (5.29,5.2) {$\accN$};
  \node [ttext](ib2) at (4.22,4.7) {$\ib$};
  \node [ttext](cc2) at (4.55,4.7) {$\cc$};
  \node [ttext](cuf2) at (4.88,4.7) {$\cuf$};
  \node [ttext](or2)  at (5.29,4.7) {$\outr$};
  \draw [->,dashed] (s1) -- (ib1);
  \draw [->,dashed] (s2) -- (ib1);
  \path [->,bend right,dashed] (s3) edge (ib3);
  \path [->,bend right] (or1) edge (ib2);
  \path [->, out=230, in=160] (or2) edge (ib3);
  \draw [->,shorten <= -0.06cm] (or3) -- (out3);
  \draw [->,shorten <= -0.06cm] (or3) -- (out2);
  \draw [->,shorten <= -0.06cm] (or2) -- (out1);
\end{tikzpicture} %
}
  \caption{An \amcs with three contexts, three sensors on the left side, and three output streams on the right side. 
          A solid line represents a flow of information from a context to its stakeholder streams, whereas a dashed line 
          indicates sensor data written to the input buffer of a context.}\label{fig:amcs}
\vspace{-8pt}
\end{figure}

We continue with a big picture on the mode of operation of an \amcs.
Every context in an \amcs asynchronously decides whether the data that is currently available on
its input information buffer \ib is sufficient to start a computation using the current knowledge base \kb of the context.
That is done by the \define{computation controller} \cc of the context, formally a binary relation that holds all pairs of input buffers and knowledge bases for which a computation should begin.
When the computation controller decides to start a computation, 
first, the context is updated using the \define{context update function} \cuf:
it maps the input buffer \ib (as it was when the computation has started) and the current knowledge base \kb of a context $C$ to a new \define{configuration} of $C$,
where a configuration of $C$ comprises the knowledge base \kb, the semantics \accN,
and the whole \define{context management} of $C$, that is,
the computation controller \cc,
set of \define{output rules} \outr (see below), and
the context update function \cuf itself.
That means, except for its unique name and the underlying logic suite, a context can change completely after every computation.
Then, one-by-one, the acceptable belief sets of the updated knowledge base are computed according to the updated semantics \accN.
Each of the computed belief sets is matched against the set of output rules of the context. A belief set can activate a rule and thereby determine which information should be sent to which stakeholders of the context (these can be other contexts or output streams of the \amcs).
A computation finishes after the final acceptable belief set has been computed.
Then, notifications about the end of the computation are sent to the stakeholders.
Figure~\ref{fig:amcs} depicts an \amcs with three contexts.

Preparing the data in the input information buffer of a context for processing is the focus of this paper.
Therefore, we now have a closer look at this data.
As stated in Definition~\ref{def:data} it consists of data sets.
These packages can come from two different sources: a data set $\data=\tuple{\name,I}$ can origin from a sensor named $\name$, in which case the \amcs framework simply assumes the package appears at some time point in the buffer and $I$ contains the corresponding sensor readings; in the other case, the package is sent from another context. Then, $\name$ is the name of this context $C$ and $I$ contains either the dedicated symbol $\eoc\in\IL$ that notifies that $C$ has finished a computation or pieces of information generated by the output rules of $C$:
an \define{output rule} $r$ for context $C= \tuple{\name,\ls}$ is an expression of the form \begin{align}
  \label{outputrule}
    \tuple{\name',\info} \la& b_1,\ldots, b_j,\naf\ b_{j+1},\ldots,\naf\ b_m,
\end{align}
such that $\name'\in\names$ is the name of a context or an output stream, $\info\in\IL$ is a piece of information,
and every $b_\ell$ ($1 \leq \ell \leq m$) is a belief for $C$, \iec $b_\ell\in \bs$ for some $\bs\in\setbs$ where \setbs is the set of possible belief sets of \ls.
We call $\name'$ the \define{stakeholder} of $r$, \tuple{\name',\info} the head of $r$ denoted by \head{r}, and $b_1,\ldots, b_j,$ $\naf\ b_{j+1},\ldots,\naf\ b_m$ the body \body{r} of $r$.
Moreover, we say that $r$ is active under \bs, denoted by $\bs\models\body{r}$, if 
$\{b_1,\ldots, b_j\}\subseteq\bs$ and $\{b_{j+1},\ldots, b_m\}\cap\bs=\emptyset$.
Intuitively, the stakeholder is a reference to the addressee of information \info
and when an output rule becomes active, the information is sent to the stakeholder, \iec the input stream of another context or an output stream of the \amcs.
The \define{output} of $C$ with respect to a set \outr of output rules for $C$ under a belief set $\bs\in\setbs$ relevant for $\name'$
is the data set
$$
\relout{C}{\bs}{\outr}{\name'}=\tuple{\name,\{i\mid r\in\outr, \head{r}=\tuple{\name',\info}, \bs\models\body{r}\}}.
$$
So, intuitively, if $\name'$ refers to some context $C'$ and $C$ has computed belief set $\bs$, then
$\relout{C}{\bs}{\outr}{\name'}$ will be put on the information buffer that represents the input stream of $C'$. Note that the first component of the data set is the name \name,  serving as reference to context $C$ as the source of information. To summarise, the head of an output rule determines the name $\name'$ of target and the name $\name$ in a data set points to the source of the information.

\begin{figure}[t]
   \centering
   \resizebox{0.5\columnwidth}{!}{%
\begin{tikzpicture}
  \tikzstyle{context}=[rounded corners]
  \tikzstyle{senI}=[fill=black]
  \tikzstyle{bnode}=[rectangle,minimum width=2cm,minimum height=1cm]
  \tikzstyle{sonode}=[rectangle,minimum width=1.07cm, minimum height=1.4cm, font=\tiny,above,yshift=0.11cm]

  \draw [context] (3,0) rectangle node (nav) [bnode] {Navigation} (5,1);
  \draw [context] (0,2) rectangle node (rum) [bnode]{Amb Manager} (2,3);
  \draw [context] (3,2) rectangle node (tp) [bnode] {Task Planner} (5,3);
  \draw [context] (3,4) rectangle node (ca) [bnode] {Case Analyser} (5,5);
  \draw [context] (0,4) rectangle node (mdb) [bnode]{Med Ontology} (2,5);

  \draw [senI] (6.24,4.75) circle (0.05);
  \draw (6.24,4.75) circle (0.2) node (seme){};
  
  \draw (5.88,4) rectangle (6.12,4.24) node [yshift=-0.12cm] (oeme) {};
  \draw (6.12,4.24) -- (6.75,4.24);
  \draw (6.12,4) -- (6.5,4);

  \draw [context, dotted] (5.68,3.8) rectangle node [sonode] {ER Employee}(6.95,5.2);
  \draw [senI] (6.24,2.75) circle (0.05) node (scd) {};
  \draw (6.24,2.75) circle (0.2);
  
  \draw (5.88,2) rectangle (6.12,2.24) node [yshift=-0.12cm] (ocd) {};
  \draw (6.12,2.24) -- (6.75,2.24);
  \draw (6.12,2) -- (6.5,2);

  \draw [context, dotted] (5.68,1.8) rectangle node [sonode] {Case Dispatcher}(6.95,3.2);
  \draw [senI] (6.24,0.75) circle (0.05);
  \draw (6.24,0.75) circle (0.2) node[label={[label distance=1,font=\tiny]Traffic state}] (sts) {};
  \draw [senI] (-1.24,0.75) circle (0.05) node (samb) {};
  \draw (-1.24,0.75) circle (0.2);
  
  \draw (-0.88,0) rectangle (-1.12,0.24) node [yshift=-0.16cm] (oamb) {};
  \draw (-1.12,0.24) -- (-1.75,0.24);
  \draw (-1.12,0) -- (-1.5,0);

  \draw [context, dotted] (-0.68,-0.2) rectangle node [sonode] {Ambulance}(-1.95,1.2);

  \path [->] (rum) edge (tp)
  (tp.250) edge (nav.110)
  (nav.70) edge (tp.290)
  (ca) edge (tp)
  (ca.170) edge (mdb.10)
  (mdb.350) edge (ca.190);

  \path [->,dashed]
  (seme) edge (ca)
  (ca) edge[solid] (oeme)
  (scd) edge (tp)
  (tp) edge[solid] (ocd)
  (sts) edge (nav)
  (samb) edge (rum.south)
  (tp.195) edge[solid] (oamb);
  \draw (-0.5,-0.5) rectangle (5.5,5.5);
  \draw (3.5,5.25) rectangle node [font=\tiny] {CAET Management} (5.5,5.5);
\end{tikzpicture}
}
   \caption{An \amcs for Computer-Aided Emergency Team Management}
   \label{fig:caetm}
\vspace{-8pt}
\end{figure}
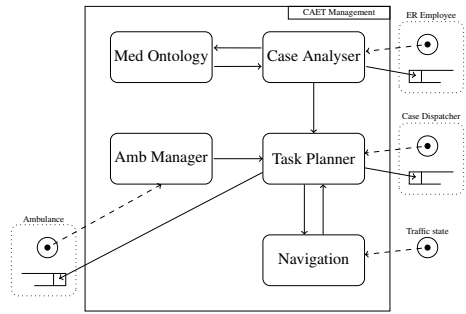

As a running example, we reuse a scenario from previous work~\cite{EllmauthalerP14}
dealing with the coordination and handling of assignments of medical rescue units.
An \amcs consisting of five context is used to model the scenario. It is depicted in Figure~\ref{fig:caetm}.
The system gives assistance during the rescue call, helps in assigning priorities and rescue units to a case, and 
assists in the necessary communication among all involved parties.

In our examples we will deal with the following contexts:
\begin{description}
\item[Case Analyser]
This context implements a computer-aided call handling system which assists an emergency response employee during answering an emergency call.
It chooses which questions need to be asked based on previous answers and checks whether answers are inconsistent (\egc amniotic sac bursts when the gender is male).
For these purposes the case analyser context may also consult a medical ontology represented by another context.
The communication with the ER~employee is represented, on the one hand, as a sensor that reads the input of the employee and, on the other hand, by an output stream which prints the questions and results on a computer screen.
During the collection of all the important facts for this emergency case, the analyser computes the priority of the case and passes it to the task planner.
\item[Task Planner]
This context keeps track of emergency cases.
Based on the priority and age of a case and the availability and position of ambulances it suggests an efficient plan of action for the ambulances to the (human) case dispatcher.
The dispatcher may approve some of the suggestions or all of them.
If the dispatcher has no faith in the given plan of action, she can also alter it at will.
These decisions are reported back to the planning system such that it can react to the alterations and provide further suggestions.
Based on the final plan, the task planner informs the ambulance about their new mission.
\item[Amb Manager]
The ambulance manager is a database, which keeps track of the status and location of ambulance units.
Each ambulance team reports its status (\egc to be on duty, waiting for new mission, \ldots) to the database (modelled by the sensor ``Ambulance'').
Additionally, the car periodically sends GPS-coordinates to the database.
These updates will be pushed to the task planner.
\end{description}

\section{Declarative Packing of Data Sets}\label{sec:main}%

Unlike previous proposals for heterogeneous multi-context systems, in the static as well as the dynamic setting~\cite{BrewkaE07,BrewkaEFW11,Brewka2014,GKL2014}, 
\amcss do not use a synchronised equilibria semantics. 
For applications that do not require the tight semantic integration offered by equilibria,
\amcss have the advantage that each context can run at its own pace and that the overall computational complexity is in general significantly lower.
A natural consequence of asynchronicity is that we cannot use bridge rules that are typically used in multi-context systems because they depend on synchronised belief sets (so-called belief states).
Instead, output rules are used that are defined by the context that sends information rather than the context that receives it.
As a consequence, the receiving context is not in control of what data it receives on its input buffer and in what order the data arrives.
We propose to declaratively specify \define{packages of data sets} that the context accepts as input. 
The goal is to allow the system to automatically pack incoming data in a form suitable as input for the logic of the context. At the same time, the method gives the context control over the data it receives.

 \begin{figure}[t]
   \centering
   \includegraphics[width=.8\textwidth]{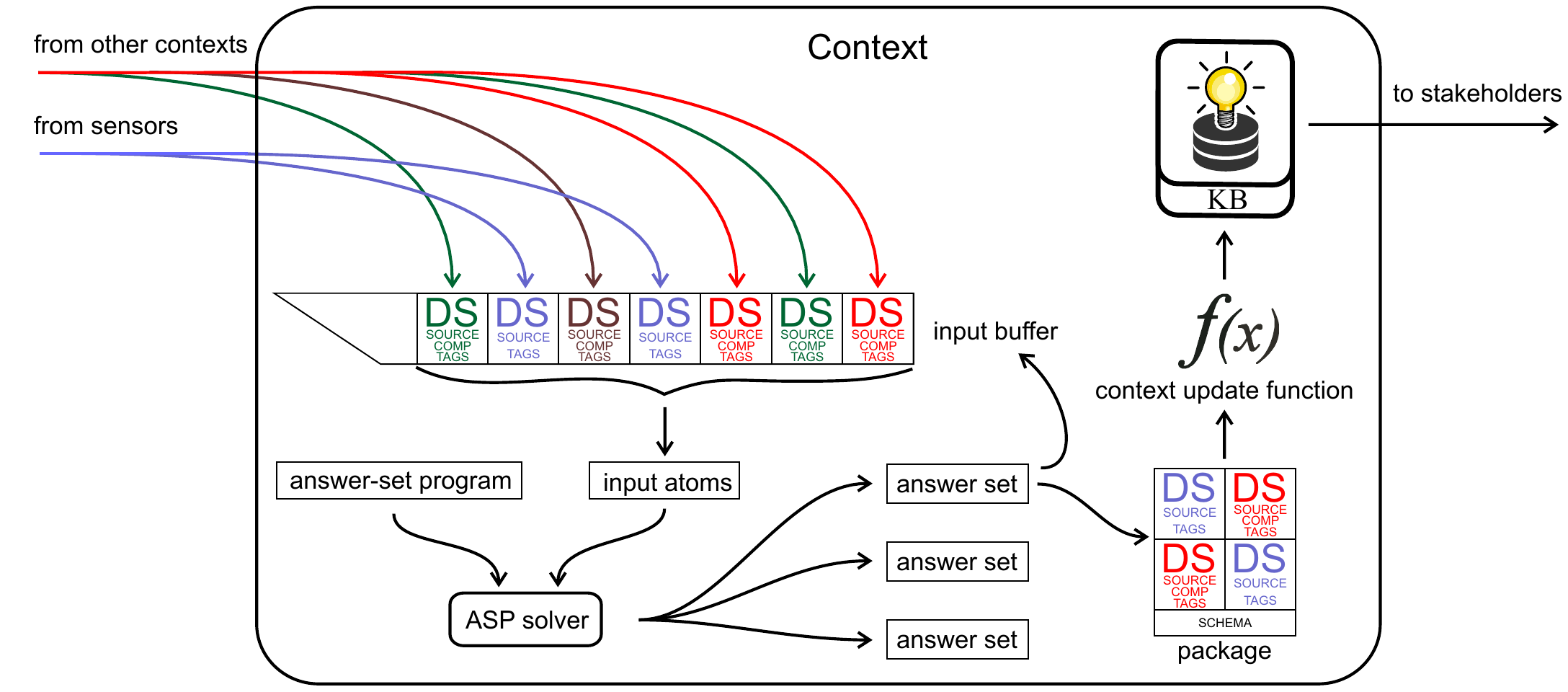}
   \caption{Overview of the approach. Data sets from other contexts and sensors are written into the input buffer of a context. Different colours of data sets mean that they are from a different source. The data in the input buffer is represented as input atoms to an answer-set program. One of the resulting answer sets is chosen. It contains directives on how to change the data on the input buffer as well as how to pack data sets. Resulting packages are associate with a package schema. Based on the schema and the chosen data sets, the context update functions determines changes of the knowledge base, the semantics to use, as well as changes in the output rules of a context. After evaluation, the context sends it output to its stakeholders.}
   \label{fig:packing}
\vspace{-8pt}
 \end{figure}

As means of specification we propose to rely on answer-set programming.
It offers a rich language for representing complicated problems and efficient ASP solvers are available.
By using ASP, we profit from its elaboration tolerance (small changes in the requirements typically require only minimal changes of the program)
and its elegant and concise declarative specifications.
Regarding computational complexity, the packing and filtering tasks we deal with will in many cases not require the full expressive power of ASP from a complexity theoretic point of view.
Often, very efficient fragments of ASP (such as programs with stratified negation) will suffice.
Nevertheless, there can be cases where it is handy to exploit the capability of ASP to solve problems from NP and beyond, e.g., when an \amcs combines pre-existing contexts that are not expressive enough to solve a required task.
Additionally, ASP offers different optimisation frameworks that we also want to exploit in our approach.

An answer-set program is a set of logic-programming rules of the form
\begin{equation}
a \leftarrow b_{1}, \ldots, b_{m}, \naf\ b_{m+1}, \ldots, \naf\ b_{n}\ ,
\end{equation}
where $a$, and every $b_{i}$ for $1\leq i\leq n$ are first order atoms and \naf denotes \emph{default negation}. The intuition is that if one knows that all atoms $b_{1}, \ldots, b_{m}$ are true and there is no evidence that some $b_{m+1}, \ldots, b_{n}$ is true, than one can derive that $a$ is true. 
The overall semantics of an answer-set program is given by its answer sets: dedicated models of the program that satisfy the implications of the program rules and adhere to certain minimality conditions.
For a formal account of (different) ASP semantics we refer to the vast body of literature on ASP (starting points are \cite{Baral2003,Lifschitz08b}).
We assume familiarity with the ASP language of the \clingo solver~\cite{gekakaosscsc11a,GebserKK0S15}
that supports aggregates, function symbols, user-defined functions, and further popular language extensions\footnote{We will make uses of \prolog style lists which are not part of the \clingo language but can easily be simulated by function symbols. As the simulation relies on nested functions it leads to heavy use of bracketing which we avoid for the sake of readability.\label{foot:list}}.

The idea is that the \amcs engine provides information about the data sets on a context's input stream in form of ASP facts.
They are the input of an answer-set program that decides 
\bi
\item if the data in the buffer is sufficient for passing it on for processing (\iec to the update function and subsequently to the knowledge base);
\item which data sets should form a package that is passed on for processing; and
\item which data sets should be deleted from or remain on the input buffer.
\ei
These decisions are provided by directives encoded in atoms in the resulting answer sets.
When and how often and the ASP evaluation should take place depends on the needs of the application.
One could, \egc re-evaluate whenever new data arrives on the input buffer or use fixed time intervals to reduce computation efforts.
Regarding the components of an \amcs, the answer-set program can be seen as an implementation of the computation controller \cc  of a context
as it decides when to start a computation.
Moreover, it partially implements the context update function \cuf, as it can remove from or leave data set on the input buffer.
The packages of data sets generated  by the answer-set program are input to the components of the \amcs engine that implements the remaining functionality of the update function.
The overall method is illustrated in Figure~\ref{fig:packing}.

\begin{example}\label{ex:basic}
Consider an \amcs for the emergency team management scenario as in Figure~\ref{fig:caetm}.
The task planner context receives information about available ambulances from the ambulance manager context
and emergency cases to assign from the case analyser.
For now, we assume that every data set from the case manager represents one case that needs assignment.
A sensible input for the task planner's knowledge base consists of exactly one case that needs assignment (\egc the one with the highest priority---\cf Section~\ref{sec:opt}) and all information about currently available ambulances.
A representation of a possible input buffer could contain the following facts
\begin{footnotesize}
\begin{verbatim}
ds_avail(ca_ds11). ds_avail(ca_ds12).
ds_avail(am_ds54). ds_avail(am_ds55). ds_avail(am_ds56).
source(ca_ds11,ctxt_case_anl). source(ca_ds12,ctxt_case_anl).
source(am_ds54,ctxt_amb_mng). source(am_ds55,ctxt_amb_mng).
source(am_ds56,ctxt_amb_mng).
\end{verbatim}
\end{footnotesize}
stating which data sets are available and what their source is, \egc
\verb|ds_avail(ca_ds11)| encodes that a data set identified by the constant \verb|ca_ds11| is
available in the buffer and \verb|source(ca_ds11,ctxt_case_anl)| that the data set stems from the case analyser context.

The following answer-set program can be used for packing.
\begin{lstlisting}
aux_case_avail :- ds_avail(DS),source(DS,ctxt_case_anl).
aux_ambulance_avail :- ds_avail(DS),source(DS,ctxt_amb_mng).
process_as_schema(sch1) §\label{line:procAs}§:- aux_case_avail, aux_ambulance_avail. 
in_pack(DS) :- ds_avail(DS), source(DS,ctxt_amb_mng).
1 {aux_case_in_pack(DS) §\label{line:cardinality}§: ds_avail(DS), source(DS,ctxt_case_anl)} 1. 
in_pack(DS) §\label{line:inPackAux}§:- aux_case_in_pack(DS). 
rm_pack.
\end{lstlisting}
The first two rules derive auxiliary atoms that indicate whether data sets from the case analyser, respectively, the ambulance manager are available.
If both is the case the rule in line~\ref{line:procAs} derives the directive \verb|process_as_schema(sch1)| which means that the selected data sets form a package of schema
\verb|sch1| and should be passed on for processing. A schema can be seen as the type of the package that may influence how the context processes the package.
The selection of data sets is expressed by atoms of the unary predicate \verb|in_pack|: the fourth rule adds all data sets that come from the ambulance manager
to the package; the rule in line~\ref{line:cardinality} is a fact enforcing a cardinality constraint. 
It expresses that exactly one data set from the case analyser, thus the information about a single open case, should be part of the package.
The rule in line~\ref{line:procAs} derives a corresponding \verb|in_pack| directive atom for the \verb|aux_case_in_pack| atom.
Finally, the directive \verb|rm_pack| removes all data sets from the input buffer that are part of the chosen package.

Together with the input facts the program has two answer sets,
one containing the directives
\begin{footnotesize}
\begin{verbatim}
in_pack(ca_ds12), in_pack(am_ds54), in_pack(am_ds55),
in_pack(am_ds56), process_as_schema(sch1), rm_pack
\end{verbatim}
\end{footnotesize}
and the other one 
the same but {\footnotesize \verb|in_pack(ca_ds11)|}  instead of {\footnotesize \verb|in_pack(ca_ds12)|}.
\end{example}

An overview over all input and directive atoms is given in Table~\ref{tab:atoms}.
As illustrated by Example~\ref{ex:basic}, we propose to represent data sets on the object level of the ASP language.
The same we can do with computations, where the constants  that identify individual data sets and computations are generated at runtime by the \amcs engine.
The identifiers for contexts or input streams are their names from \names.

\begin{table}[t!]
\begin{footnotesize}
\begin{tabularx}{\textwidth}{|l|c|X|}
\hline
\emph{\bf Atom}&\emph{\bf Type}&\emph{\bf Description}\\
\hline
\verb|ds_avail(ds)| & Input & Data set \verb|ds| is available on the input buffer\\
\verb|ds_comp(ds,comp)| & Input & Data set \verb|ds| belongs to computation \verb|comp|\\
\verb|source(comp,ctxt)| & Input & Computation \verb|comp| is a computation of context \verb|ctxt|\\
\verb|source(ds,ctxt)| & Input & Data set \verb|ds| originates from context \verb|ctxt|\\
\verb|source(ds,therm)| & Input & Data set \verb|ds| originates from sensor \verb|therm|\\
\verb|eoc(comp)| & Input & Computations \verb|comp| has ended.\\
\verb|tag(comp,solves(probl1))| & Input & Computations \verb|comp| is tagged with function \verb|solves(probl1)|\\
\verb|tag(ds,"optimum")| & Input & Data set \verb|ds| is tagged with string \verb|"optimum"|\\
\verb|time(1000)| & Input & An external clock provides \verb|1000| as current time\\
\verb|ignore(comp)| & Directive & Ignore future data sets of computation \verb|comp|\\
\verb|add_tag(comp,best(3))| & Directive & Tag computation \verb|comp| with function \verb|best(3)|\\
\verb|rm_tag(comp,"trusted")| & Directive & Remove the tag \verb|"trusted"| from computation \verb|comp|\\
\verb|rm(comp)| & Directive & Remove all data sets of computation \verb|comp| from the input buffer\\
\verb|rm(ds)| & Directive & Remove data set \verb|ds| from the input buffer\\
\verb|rm_pack| & Directive & Remove all data sets of processed packages from the input buffer\\
\hline
\multicolumn{3}{|c|}{Variant: one package per answer set}\\
\hline
\verb|in_pack(ds)| & Directive & Data set \verb|ds| is considered part of the package\\
\verb|process_as_schema(sch)| & Directive & The data sets defined by \verb|in_pack/1| atoms form a package of schema \verb|sch| and are passed on for processing\\
\hline
\multicolumn{3}{|c|}{Variant: multiple packages per answer set}\\
\hline
\verb|process(sch,[ds1,ds3,ds7])| & Directive & The data sets in the list \verb|[ds1,ds3,ds7]| form a package of schema \verb|sch| and are passed on for processing\\
\hline
\end{tabularx}
\end{footnotesize}
\caption{Input and directive atoms for package specifications. Atoms of other predicates can be used as needed and are considered auxiliary.}\label{tab:atoms}
\end{table}

\begin{example}\label{ex:comp}
Remember that the case analyser is a context responsible for the collection of data about an emergency case during a phone call between someone in need and a human employee.
This collection can be a lengthy process and the data available about the case can evolve over time.
Hence, let us now assume that the case analyser provides not only a single data set for one case but multiple ones such 
that the whole handling of a case is considered a computation that has multiple acceptable belief states representing different states of refinement of a case.
Thus, we can have multiple data sets for the same computation of the case analyser in the input buffer of the task planner.
Moreover, due to the asynchronous nature of \amcss, it can happen that there are data sets of multiple computations available at the same time.
Therefore, we need to be able to distinguish to which computation a data set belongs to.
An input for packing could then consist of the facts given next.
\begin{footnotesize}
\begin{verbatim}
ds_avail(ca_ds21). ds_avail(ca_ds22).
ds_avail(ca_ds24). ds_avail(ca_ds25).
ds_comp(ca_ds21,ca_comp35). ds_comp(ca_ds22,ca_comp35).
ds_comp(ca_ds24,ca_comp36). ds_comp(ca_ds25,ca_comp36).
ds_avail(am_ds54). ds_comp(am_ds54,am_comp61).
eoc(ca_comp35). eoc(am_comp61).
source(ca_ds21,ctxt_case_anl). source(ca_ds22,ctxt_case_anl).
source(ca_ds24,ctxt_case_anl). source(ca_ds25,ctxt_case_anl).
source(ca_comp35,ctxt_case_anl). source(ca_comp36,ctxt_case_anl).
\end{verbatim}
\end{footnotesize}
Compared to Example~\ref{ex:basic}, the additional binary predicate \verb|ds_comp| states to which computation a  data set belongs
and the use of \verb|source| atoms is extended to computations.
The \verb|eoc| atoms state which computations have finished (this is useful if the context needs to wait for all data sets of a computation).
In our setting, for every emergency case it is always the latest data set provided by the case analyser that we want to consider
in the task planner. 
For identifying these in the answer-set program we can use a comparison relation that indicates the order of arrival of data sets.
In this example, we assume that the constants representing data sets are assigned by system in a way that the internal term comparison relation of \clingo
respects the arrival order. Alternatively, the system could provide the arrival order by further input facts.
We can adapt the previous packing program as follows.
\begin{lstlisting}
aux_case_avail :- ds_avail(DS),source(DS,ctxt_case_anl).
aux_ambulance_avail :- ds_avail(DS),source(DS,ctxt_amb_mng).
process_as_schema(sch1) :- aux_case_avail, aux_ambulance_avail.
in_pack(DS) :- ds_avail(DS), source(DS,ctxt_amb_mng).
1 {aux_selected_case_comp(CO) §\label{line:selComp}§: source(CO,ctxt_case_anl),ds_comp(DS,CO)} 1. 
in_pack(MDS) :- MDS = #max{DS : ds_avail(DS), ds_comp(DS,CO), aux_selected_case_comp(CO)}.
rm_pack.
\end{lstlisting}
The rule in line~\ref{line:selComp} selects one of the computations from the case analyser context for which a data set is available.
The next rule selects the latest data set from this computation (that with the maximum identifier according to the term comparison relation of \clingo) to be part of the package.
We get two answer sets, one including  \verb|in_pack(ca_ds22)| representing the latest state of the case covered in computation \verb|ca_comp35| and the other one
directive \verb|in_pack(ca_ds25)| selecting the latest data set \verb|ca_ds25| of computation \verb|ca_comp36|.
\end{example}

\subsection{Tagging}
Both examples showed how we can select a data set to be part of a package based on meta information such as the number of available data sets or their order.
That is, we did not use any information about the content or purpose of the data set itself.
To gain flexible control for the selection we propose to use  a \define{tagging system} for data sets and computations.
Every data set and every computation is associated with a set of arbitrary (ground) \clingo terms that serve as tags.

\begin{example}\label{ex:tags}
In Example~\ref{ex:comp}, the latest data set of a computation that is available in the buffer is considered to describe the corresponding emergency case.
This was based on the assumption that the consecutive data packages of a computation matches the evolution of data available for the case.
By design of \amcss, one context may only have one computation at a time.
That means that, if we modelled the \amcs as in the previous example, 
if we first deal with a case $a$ and then a new case $b$ is handled by the case analyser,
the previous computation has finished and we could not continue to refine data on case $a$.
In order to allow such late refinements, 
we drop the assumption that a single computation of the case analyser contains all data sets to a case.
Instead, we tag every data set with an uninterpreted function symbol \verb|case(id,i)|, where \verb|id| is an identifier for the case and \verb|i| is the index of the current revision of the case data.
In the new setting we could have the following input facts.
\begin{footnotesize}
\begin{verbatim}
ds_avail(ca_ds26). ds_avail(ca_ds27). ds_avail(ca_ds28).
ds_comp(ca_ds26,comp37). ds_comp(ca_ds27,comp38).
ds_comp(ca_ds28,comp39).
source(ca_ds26,ctxt_case_anl). source(ca_ds27,ctxt_case_anl).
source(ca_ds28,ctxt_case_anl). source(comp37,ctxt_case_anl).
source(comp38,ctxt_case_anl). source(comp39,ctxt_case_anl).
tag(ca_ds26,case(c1,1)).tag(ca_ds27,case(c2,1)).
tag(ca_ds28,case(c1,2)).
\end{verbatim}
\end{footnotesize}
There are three data sets from the case analyser, each being the result of a different computation.
Yet, data sets \verb|ca_ds26| and \verb|ca_ds28| deal with the same emergency case  as indicated by the tags.
As \verb|ca_ds28| has the higher index, this data set should be selected if this case is processed.
We can also use tags for differentiating types of data sets. The following packing program distinguishes
between data sets from the ambulance manager that state that an ambulance is available and data sets indicating 
a broken ambulance.
\begin{lstlisting}
aux_case_avail :- ds_avail(DS),source(DS,ctxt_case_anl).
aux_ambulance_avail :- ds_avail(DS),source(DS,ctxt_amb_mng), tag(DS,"available"), not aux_some_amb_broken.
aux_some_amb_broken :- ds_avail(DS),source(DS,ctxt_amb_mng), tag(DS,"broken").
process_as_schema(sch1) :- aux_case_avail, aux_ambulance_avail, not aux_some_amb_broken.
process_as_schema(sch2) :- aux_some_amb_broken.
in_pack(DS) :- ds_avail(DS),source(DS,ctxt_amb_mng), not aux_some_amb_broken.
in_pack(DS) :- ds_avail(DS),source(DS,ctxt_amb_mng), tag(DS,"broken").
1 {aux_selected_case(C) §\label{line:selCase}§: tag(DS,case(C,I)),ds_avail(DS), source(DS,ctxt_case_anl)} 1 :- not aux_some_amb_broken.
in_pack(MDS) §\label{line:maxIndex}§:- MI = #max{I:tag(DS,case(C,I)),ds_avail(DS),source(DS,ctxt_case_anl)}, tag(MDS,case(C,MI)), ds_avail(MDS), aux_selected_case(C), source(MDS,ctxt_case_anl).
rm_pack.
rm(DS) §\label{line:rmDS}§:- ds_avail(DS), source(DS,ctxt_case_anl), aux_selected_case(C), tag(DS,case(C,I)), not in_pack(DS).
\end{lstlisting}
The rule in line~\ref{line:selCase} derives one auxiliary atom indicating which case the answer set deals with.
The next rule then adds the data set dealing with this case that has the highest index to the package.

Unlike the previous examples, the program defines packages of two different schemata: schema \verb|sch1| contains
one data set defining a case and all data sets about available rescue units. 
But as soon as some ambulance is broken, only packages of schema \verb|sch2| are processed. These contain all data sets
about broken ambulances but none about cases.

Finally, the example illustrates another feature for removing data sets from the buffer: the final rule ensures that 
data sets dealing with the chosen case that have a non-maximal index should be removed.
Note that with the given encoding they are even removed if no package is processed:
for the given input fact, the program does not derive the \verb|process_as_schema| directive, since this time no ambulance is available.
Still, for case \verb|c1| the directive \verb|rm(ca_ds26)| for removing data set \verb|ca_ds26| is derived.
\end{example}
The example illustrated how we can use tags in the answer-set program but so far we did not address how these are assigned.
We see multiple sensible mechanisms for doing so:
\be
\item Tags can be assigned by the sending context. Formally, a function is added to the context that, given the current configuration of the context, assigns tags to the data sets and computations it produces. By taking the current configuration of the context into account, the tags can also provide relevant meta data. For example, when a context solves an optimisation problem, a tag can indicate whether a data set is derived from an optimal accepted belief set.
\item It also makes sense when the receiving context can generate tags on reception of data sets in a similar way.
      This additional variant is useful, \egc in a distributed setting when we are only in control of the receiving context
      or when multiple contexts are receiving the same data sets but require different tagging.
\item An \amcs engine could itself provide tags giving access to system parameters such as the time when a data set was created. 
      The latter is interesting when we also have access to the current time in the answer-set program by a further input fact stating the current time.
      This way we can, \egc wait for a minute after receiving a data set with a non-optimal solution whether a better solution is provided. In case it is not,
      the suboptimal solution will be passed on by the first evaluation of the packing program after the time-out.
\item The answer-set program itself can be extended to modify tags for computations and data sets that remain on the input buffer.
Thus, by means of tags, the packing mechanism becomes a stateful system that can remember or forget information.
We propose two directives \verb|add_tag| and \verb|rm_tag| for this purpose. Applications include 
strategies to avoid starvation of data sets and marking computations as trusted or untrusted based on the data sets they provide over time.
In case that an ongoing computation is identified as not of interest to the context, we additionally introduce the directive \verb|ignore(comp)| 
that permanently bans further packages of computation \verb|ignore(comp)| from entering the input buffer.
\ee

\subsection{Multiple Answer Sets and Multiple Packages}\label{sec:opt}

An important principle of answer-set programming is that a program encodes a problem whose solutions correspond to the answer sets of the problem.
This is also a valid point of view for our packing programs: the answer sets we get correspond to solutions of a packing problem.
As a consequence, when a problem has multiple solutions, we also deal with multiple answer sets and, indeed, all the programs in Examples~\ref{ex:basic}-\ref{ex:tags}
have multiple answer sets.
Our proposal to deal with multiple answer sets is to let the \amcs engine compute at most one of them in the first place.
The underlying assumption is that every answer set is equally well suited as all of them are considered a solution.
Therefore, a situation where a program has answer sets that are considered inferior to others
requires modifying the program such that only wanted answer sets remain. 
ASP offers many excellent features for specifying preferences and finding optimal solutions (see~\cite{Baral2003,BrewkaD0S15,DelgrandeSTW04}).

\begin{example}\label{ex:opt}
Remember that the packing program in Example~\ref{ex:basic} has two answer sets
and that the output of the case analyser context contains information about the priority of an emergency case.
So far we did not make use of this information. 
In order to do so we assume that the case analyser assigns tags specifying the case priority to its data sets.
Then, in addition to the input given in Example~\ref{ex:basic},
we could have
\begin{footnotesize}
\begin{verbatim}
tag(ca_ds11,3). tag(ca_ds12,7).
\end{verbatim}
\end{footnotesize}
as input atoms that hold the numeric priority values of the cases handled by the respective data sets.
In order to get only one answer set dealing with case \verb|ca_ds12| that has higher priority it suffices to add
the following maximize statement.
\begin{lstlisting}
#maximize{P:aux_case_in_pack(DS),tag(DS,P)}.
\end{lstlisting}
\end{example}
\vspace{-8pt}
Another option for handling multiple answer sets is to consider them all, \iec multiple packages would be passed on for processing and all the modifications of the input buffer appearing in some answer sets would be applied.
This variant however would require very careful modelling. First, inconsistencies might occur, \egc
tags could be assigned and removed at the same time.
Moreover, overlapping of answer sets might often be a problem: 
both answer sets in Example~\ref{ex:basic} contain the same data sets about available ambulances. 
That is, the same ambulance could subsequently be assigned for both cases.
As ASP offers little means for reasoning across answer sets (an exception are optimisation features as in Example~\ref{ex:opt})
such situations are hard to avoid or lead to complicated encodings.

If one wants to process multiple packages at a time, we suggest to encode all desired packages in one answer set using lists (or function symbols as stated in footnote~\ref{foot:list}).
To this end, we propose to replace directives \verb|in_pack| and \verb|process_as_schema| by a single new one.
The binary directive \verb|process| takes as first argument the schema, similar as the argument of \verb|process_as_schema|.
The second argument is a list containing the data sets that form one of the packages to be processed.
Deriving multiple \verb|process| atoms we can create multiple packages.
\begin{example}\label{ex:list}
The following code divides the data sets about available ambulances in three disjunct packages.
\begin{lstlisting}
aux_min_amb_ds(MDS) :- MDS = #min{DS : ds_avail(DS),source(DS,ctxt_amb_mng)}.
aux_amb_in_between(DS1,DS2) :- ds_avail(DS1),source(DS1,ctxt_amb_mng), ds_avail(DS),source(DS,ctxt_amb_mng), ds_avail(DS2),source(DS2,ctxt_amb_mng), DS1<DS, DS<DS2.
aux_amb_ds_nr(MDS,0) :- aux_min_amb_ds(MDS).
aux_amb_ds_nr(DS2,I+1) :- aux_amb_ds_nr(DS1,I), ds_avail(DS2),source(DS2,ctxt_amb_mng), DS1<DS2, not aux_amb_in_between(DS1,DS2).
aux_build_package(I,I,DS) §\label{line:aux_build1}§:- I=0..2,aux_amb_ds_nr(DS,I). 
aux_build_package(I\3,I,[DS|T]) §\label{line:aux_build2}§:- aux_build_package(I\3,I-3,T),aux_amb_ds_nr(DS,I).
package(sch,L) :- aux_build_package(M,MI,L),M=0..2,MI=#max{I:aux_build_package(M,I,_)}.
\end{lstlisting}
The first four rules assign every data set an index according to \clingo's built in order
using the auxiliary binary predicate \verb|aux_amb_ds_nr|.
The next two rules create partial lists, where the first argument of the \verb|aux_build_package| atoms is the number of the final list (always $0$,$1$, or $2$),
the second argument is the index of the data set added to the list, and the third argument is the partial list. Note that \verb|I\3| stands for \verb|I| modulo \verb|3|.
Finally, the last rule derives the \verb|package| directive with the maximum partial list with first argument $0$,$1$, and $2$, respectively.
Consider the following input facts.
\begin{footnotesize}
\begin{verbatim}
ds_avail(ca_ds26). ds_avail(ca_ds27). ds_avail(ca_ds28).
ds_comp(ca_ds26,comp37). ds_comp(ca_ds27,comp38).
ds_comp(ca_ds28,comp39).
source(ca_ds26,ctxt_case_anl). source(ca_ds27,ctxt_case_anl).
source(ca_ds28,ctxt_case_anl). source(comp37,ctxt_case_anl).
source(comp38,ctxt_case_anl). source(comp39,ctxt_case_anl).
tag(ca_ds26,case(c1,1)). tag(ca_ds27,case(c2,1)).
tag(ca_ds28,case(c1,2)).
\end{verbatim}
\end{footnotesize}
As a result, we get an answer set with the following directives with the desired packages.
\begin{footnotesize}
\begin{verbatim}
package(sch, [am_ds56, am_ds49, am_ds34]),
package(sch, [am_ds74, am_ds53, am_ds45]),
package(sch, [am_ds84, am_ds55, am_ds46, am_ds24]).
\end{verbatim}
\end{footnotesize}
\end{example}
\vspace{-7pt}
\section{Conclusion}\label{sec:conclusion} %
While the method presented in this paper was developed for use in the \amcs framework, it addresses a general problem
that will often occur when dealing with stream data: 
deciding when and what part of buffered incoming data to process or what information to ignore.
It was our goal in this paper to show in a range of examples that declarative specifications using ASP can be of great help here.
The rich language of ASP allows for concisely expressing how data should be bundled in packages that meet the input requirements for further processing which can be seen as a combined classification and configuration task. 
Using ASP for configuration purposes has a long tradition~\cite{SoininenNTS01,aspcud11}.
We use ASP for pre-processing and structuring of stream data
which can itself be seen as an instance of stream reasoning using ASP~\cite{DoLL11,GebserGKOSS12,NicklesM14}.
An interesting question is to what extent reasoning should take place in the packing layer or in the contexts.
The answer seems to depend on the application: reasoning should take place in the contexts if the context formalism is expressive enough and the reasoning task conceptually belongs to the context;
otherwise we can exploit ASP on the packing layer to solve problems that are too complex for the contexts or that do not fit the conceptual scope of any context.
Our method does not fully inspect the contents of data sets but relies on clearly structured meta information and tags provided by the system or the source of information.
This is a difference to data stream mining in the areas of data mining and machine learning,
dealing with approximate classification and clustering of stream data~\cite{HuangYK15,HahslerD11}.

In ongoing work, we are developing an \amcs engine that implements the features we described.
In order to guarantee good interoperability the software will make use of the efficient ASP solver \clingo as a library.

\end{document}